\documentclass[structabstract]{aa}  
\usepackage{graphicx}
\usepackage{txfonts}
\usepackage{natbib}

\begin{document}

   \title{Analytic approximate seismology of transversely oscillating coronal loops}

   \author{M. Goossens
          \inst{1}
          \and
          I. Arregui\inst{2}
	  \and
	  J. L. Ballester\inst{2}
	  \and
	  T. J. Wang\inst{3}
          }

   \institute{Centre Plasma Astrophysics, Katholieke Universiteit
                 Leuven, Leuven, B-3001, Belgium\\
                 \email{marcel.goossens@wis.kuleuven.be}
         \and
                  Departament de F\'{\i}sica, Universitat de les Illes Balears
	          E-07122 Palma de Mallorca, Spain\\
                  \email{[inigo.arregui,dfsjlb0]@uib.es}
         \and
	          Department of Physics, The Catholic University of America and NASA 
		  Goddard Space Flight Center, Code 671, Greenbelt, 
		  MD 20771, United States\\
	          \email{wangtj@helio.gsfc.nasa.gov}
             }

   \date{Received / Accepted}
   
\authorrunning{M. Goossens et al.}
\titlerunning{Analytic approximate seismology}

 
  \abstract
  {}
  {We present an analytic approximate seismic inversion scheme for damped transverse coronal 
  loop oscillations based on the thin tube and thin boundary approximation for computing the period 
  and the damping time.}
  {Asymptotic expressions for the period and damping rate are used
   to illustrate the process of seismological inversion in a simple and easy to follow manner. 
   The inversion procedure is formulated in terms of two simple functions, which are given by 
   simple closed expressions.}
  {The analytic seismic inversion shows that an infinite amount of 1-dimensional equilibrium models can reproduce the observed
 periods and damping times. It predicts a specific range of allowable values for the Alfv\'en travel  time and lower bounds for the density contrast
 and the inhomogeneity length scale. When the results of the present analytic seismic inversion are compared with those of a previous numerical 
 inversion, excellent agreement is found up to the point that the analytic seismic inversion emerges as a tool for validating results of numerical 
 inversions. Actually it helped us to identify and correct inaccuracies in a  previous numerical investigation.}
  {}


\maketitle

\section{Introduction}\label{intro}

References to coronal seismology date back to the 1980s (\citealt{REB84})
and even the 1970s (\citealt{uchida70}), but coronal
seismology remained largely a theoretical concept. This situation
changed drastically when observations from space observatories
showed that MHD waves are ubiquitous in the solar atmosphere.
Opinions might differ, but we are inclined to identify the detection
of damped transverse coronal loop oscillations in 1999 by \cite{Aschwanden99} and
\cite{Nakariakov99} in observations made with the EUV telescope on
board of the Transition Region and Coronal Explorer (TRACE) as the
real start of coronal seismology. Since then the detection of these
oscillations has been confirmed (see e.g. \citealt{SAT02}). The TRACE oscillations have periods $T$ of the order of
$\simeq 2 -10$ minutes and comparatively short damping times of the
order of $\tau_{\rm d}$ $\simeq 3 -20$ minutes. There is general
consensus that the TRACE oscillations are fast standing kink mode
oscillations. In addition, damped oscillations observed in hot coronal loops by the SUMER instrument on board SOHO
have been interpreted as standing slow mode oscillations (\citealt{Wang02,Wang03})
and the measured period has been used to determine the magnetic field strength in the loop (\citealt{Wang07}).

Theory shows that when the fast magneto-sonic kink MHD waves have
their frequencies in the Alfv\'{e}n continuum, they couple to
local resonant Alfv\'{e}n waves (\citealt{WR95,TIGO96,GAA06,Goossens08})
and get transformed into quasi-modes that are
damped by resonant absorption (\citealt{TIGO96}). This is exactly
what happens for a radially stratified cylindrical coronal loop
model (\citealt{GAA02,RR02}). Coupling of the
fast magneto-sonic MHD waves to local Alfv\'{e}n waves is a natural
phenomenon in  non-uniform coronal loops. This resonant coupling
produces a quasi-mode which in a static equilibrium is damped by
resonant absorption. The damping is independent of the dissipative
coefficients.

This finding does not mean that other damping mechanisms are ruled
out. Resonant absorption might not be the only cause of the observed
damping but it is definitely operational. Strong support for the
robust character of quasi-modes and their resonant damping comes
from a recent investigation by \cite{Terradas08} on MHD waves in
multi-stranded coronal loops. An important finding of this
investigation is that resonantly damped quasi-modes  live on
complicated multi-stranded coronal loops. They do not need nice
cylindrical magnetic surfaces as might be concluded from studies on
simplified 1-dimensional equilibrium models. The message is that
1-dimensional models are a great help to reduce the mathematical
complexity but still contain the essential physics of resonantly
damped quasi-modes.

\cite{RR02} were the first to suggest that the observed
rapid damping of the transverse oscillations of coronal loops could
be explained by resonant absorption. In the context of the heating
of solar plasmas \cite{HY88} have predicted that oscillations
in coronal loops are to undergo rapid damping. In the same context
\cite{Goossens92} derived analytical expressions for the frequency
and the damping rate of quasi-modes in static and stationary
equilibrium models.  \cite{RR02} focused on proving the
principle of resonant absorption as damping mechanism for the
transverse oscillations in coronal loops and considered one specific
numerical example.  \cite{GAA02} looked at the damping times
of 11 loop oscillation events and basically confirmed that resonant
absorption can explain the observed damping as suggested by
\cite{RR02}.

The analytic expressions derived by \citet{HY88,Goossens92,RR02},
and \cite{GAA02} for the damping rate are asymptotic in the sense 
that they are derived in the assumption that the non-uniform layer is thin. 
This is the so-called thin boundary approximation, in what follows we shall
refer to it as the TB-approximation.

The seismological studies on transverse oscillations so far are by
\cite{naka00,Nakariakov01,GAA02,Aschwanden03b}, and \cite{Arregui07}.
\cite{naka00} and \cite{Nakariakov01} used the observed
periods and theoretical estimates of the periods based on the long
wavelength or thin tube approximation (TT-approximation) for a
uniform coronal loop model to derive estimates for the magnetic field strength. 
The weak link in their analysis is the
uncertainties on the density. \cite{GAA02} used the
observed damping rates and theoretical values of the damping rates
based on the TB-approximation to derive estimates for the radial
inhomogeneity length-scale. Again, the weak link is the
uncertainties on the density. \cite{Aschwanden03b} used the
observed damping rates and the damping rates computed by
\cite{tom04b}, outside the TB-regime, to determine the density contrast. The
first study that used the observational information on both
periods and damping times in the context of resonant damping in a
consistent manner is by \cite{Arregui07}.

The important finding of \cite{Arregui07} was that an infinite amount
of 1-dimensional cylindrical  equilibrium models 
can reproduce the observed period and damping rate with the internal
Alfv\'{e}n transit time or conversely the internal Alfv\'{e}n velocity
confined to a short range. The study by \cite{Arregui07} is fully
numerical and apparently part of the physics has remained not well
understood. The objective of the present paper is to use asymptotic
expressions for the period and damping rate to illustrate in a
simple and easy to follow manner the process of seismological
inversion. The asymptotic expressions are the TT-approximation for
the period and the TB-approximation for the damping rate. When both
approximations are used simultaneously we shall refer to it as the
TTTB-approximation. We are well aware of the fact that in case of
strong damping this approximation might give inaccurate results.
However, our primary intention is to understand the process of
seismic inversion and as we shall see the asymptotic expressions
turn out to be accurate far beyond their theoretical range of
validity.


\section{Asymptotic analytic expressions for period}
The analytical expression that we shall use for the period is
obtained by (i) adopting the TT-approximation for MHD waves and by (ii) modelling the coronal loop
as a uniform cylinder  with a straight magnetic field along the
$z-$axis. The TT-approximation means that the period is
independent of the radius and that effects due to non-zero radius
are absent as far as the period is concerned. The choice of a
uniform equilibrium model means that effects due to stratification
are absent. The coronal loop is modelled as a cylindrical plasma
with constant density $\rho_{\rm i}$ embedded in an external plasma with
constant density $\rho_{\rm e}$. The coronal loop is basically a density
enhancement with $\rho_{\rm i} > \rho_{\rm e}$. The magnetic field is constant
and has the same strength both inside and outside the loop.

Our starting point is the well-known expression for the square of
the frequency of the kink mode in a uniform cylinder with a straight
magnetic field along the $z-$axis (see e.g.\ \citealt{ER83})
\begin{equation}
\omega^2 = \omega_{\rm k}^2 =  \frac {\displaystyle \rho_{\rm i}
\omega_{\rm A,i}^2 + \rho_{\rm e}  \omega_{\rm A,e}^2} {\displaystyle \rho_{\rm i} +
\rho_{\rm e}}. \label{TrueDisc1}
\end{equation}
The subscripts i and e refer to quantities respectively in the
coronal loop and in the external plasma surrounding the loop. $\omega_{\rm A}=k_{\rm z}V_{\rm A}$ is the local 
Alfv\'{e}n frequency, with $k_{\rm z}$ the parallel wavenumber, $V_{\rm A}=B/\sqrt(\mu \rho)$ the local Alfv\'{e}n 
velocity, and $B$ the magnetic field strength. 
Hence we can rewrite Eq.~(\ref{TrueDisc1}) as
\begin{equation}
\omega^2 = 2 k^{2}_{\rm z} \frac{\displaystyle B^2}{\displaystyle \mu
\rho_{\rm e}} \left(  1 + \zeta \right)^{-1}, \label{TrueDisc2}
\end{equation}

\noindent
with $\zeta = \rho_{\rm i}/\rho_{\rm e} > 1$ the density contrast.
Now we note that for the observed transverse oscillations, with a wavelength double the length $L$ of 
the loop, $k_{\rm z} = \pi/L$  and we convert frequencies to periods and rewrite Eq.~(\ref{TrueDisc2}) as
(see e.g. \citealt{Arregui08})
\begin{equation}
T  =  \tau_{\rm A,i} \; \sqrt{2} \; \; A(\zeta). \label{T}
\end{equation}

\noindent
Here $\tau_{\rm A,i}=L/V_{\rm A,i}$ is the internal Alfv\'{e}n travel time and
the function $A(\zeta)$ is defined as
\begin{equation}
A(\zeta) = \left ( \frac{\displaystyle \zeta +1}{\displaystyle
\zeta} \right )^{1/2}. \label{A}
\end{equation}

Equation (\ref{T}) is our first key equation. Let us recall that Eq.~(\ref{T}) has been
obtained by use of the TT-approximation, hence effects from non-zero radius and 
stratification are absent. This equation  expresses the
period $T$, which is an observable quantity, in terms of the
Alfv\'{e}n travel time $\tau_{\rm A,i}$ and the density contrast
$\zeta$ which are two quantities that we aim to determine from
seismic inversion. If we
have observed values  of the period $T$ and we convince ourselves
that Eq.~(\ref{T}) is a good first analytical approximation of the
period $T$ then we can invert it for either $\tau_{\rm A,i}$ or
$\zeta$. Actually, we shall do both. 
Let us first solve Eq.~(\ref{T})
for $\tau_{\rm A,i}$. Since we prefer to use dimensionless quantities
we introduce $y$ as
\begin{equation}
y = \frac{\displaystyle \tau_{\rm A,i}}{\displaystyle T}. \label{y}
\end{equation}
From Eq.~(\ref{T}) we obtain
\begin{equation}
y = \frac{\displaystyle \tau_{\rm A,i}}{\displaystyle T} =
\frac{\displaystyle 1}{\displaystyle \sqrt{2}}\; \frac{\displaystyle
1}{\displaystyle A(\zeta)} = \frac{\displaystyle 1}{\displaystyle
\sqrt{2}} \left ( \frac{\displaystyle \zeta}{\displaystyle \zeta +
1} \right )^{1/2}. \label{y1}
\end{equation}
Since we have not any information on $\zeta$, it might appear that
Eq.~(\ref{y1}) is not very helpful. However, closer inspection reveals
that it contains very valuable information. For a
given observed period $T$, Eq.~(\ref{y1}) is a parametric
representation of $y$ (or equivalently of $\tau_{\rm A,i}$) in terms
of $\zeta$. In order to stress this point we define the function
$F_1$ by use of the right hand member of Eq.~(\ref{y1}) as
\begin{equation}
F_1 \;:\;[1, \; \infty[ \;\rightarrow  \mathbf{R},\;\; \zeta \;
\Rightarrow \; F_1(\zeta) =  \frac{\displaystyle 1}{\displaystyle
\sqrt{2}} \left( \frac{\displaystyle \zeta}{\displaystyle \zeta + 1}
\right )^{1/2}. \label{F1}
\end{equation}
For later comparison in Sect.~\ref{seismology} we note that ${\mathrm d} F_1/{\mathrm d} \zeta >0$ 
and ${\mathrm d}^2 F_1 / {\mathrm d}\zeta^2 < 0$. Hence $y=F_1(\zeta)$ is a strictly increasing and concave
function of $\zeta$. We also note that
$$
F_1(1) = \frac{\displaystyle 1}{\displaystyle 2}, \;\; \lim_{\zeta
\rightarrow \infty} F_1(\zeta) = \frac{\displaystyle
1}{\displaystyle \sqrt{2}}.
$$
This means that
\begin{eqnarray}
\frac{\displaystyle 1}{\displaystyle 2} &\leq & \;y <
\frac{\displaystyle 1}{\displaystyle \sqrt{2}}, \nonumber \\
\frac{\displaystyle T}{\displaystyle 2} &\leq& \;\tau_{\rm A,i} <
\frac{\displaystyle T}{\displaystyle \sqrt{2}}. \label{y2}
\end{eqnarray}
According to inequality~(\ref{y2})  the Alfv\'{e}n travel time is
constrained to a narrow range;  whatever the density contrast is,
the Alfv\'{e}n travel time is in between $0.5 \;T$ and $0.707 \;T$.
If we accept that the density contrast $\zeta$ is not smaller than
say 1.5 then $0.5 \;T$ is replaced with $0.548\; T$ narrowing
further down the range for $\tau_{\rm A,i}$.

If we are able to determine the length $L$ of the loop, then we
can extract $V_{\rm A,i}$ and find
\begin{eqnarray}
& & V_{\rm A,i}  =  \sqrt{2} \;\; \frac{\displaystyle L}{\displaystyle
T} \;\; A(\zeta), \nonumber \\
& & \sqrt{2} \;\; \frac{\displaystyle L}{\displaystyle T} \; <
V_{\rm A,i} \leq \;\; 2 \frac{\displaystyle L}{\displaystyle T}.
\label{vAi}
\end{eqnarray}
Hence, here is the narrow range for the local Alfv\'{e}n velocity
that \cite{Arregui07} found in their numerical inversion. If the
density contrast is  not smaller than 1.5, $2\; L / T$ is replaced
with $1.825 \; L/T$ further reducing the available range for
$V_{\rm A,i}$ to less than a 25$\%$ relative margin compared  to its
possible maximal value.

With the help of the first line of inequality~(\ref{y2}) we can refine the
definition of $F_1$ and replace (\ref{F1}) with
\begin{equation}
F_1 \;:\;[1, \; \infty[ \;\rightarrow  [\frac{\displaystyle
1}{\displaystyle 2}, \;\; \frac{\displaystyle 1}{\displaystyle
\sqrt{2}} [,\;\; \zeta \; \Rightarrow \; F_1(\zeta) =
\frac{\displaystyle 1}{\displaystyle \sqrt{2}} \left(
\frac{\displaystyle \zeta}{\displaystyle \zeta + 1} \right )^{1/2}.
\label{F1a}
\end{equation}

Let us now solve Eq.~(\ref{T}) for $\zeta$ and find
\begin{equation}
\zeta = \frac{\displaystyle  2 y^2}{\displaystyle 1 -
 2 y^2}. \label{zeta1}
\end{equation}
Equation (\ref{zeta1}) is the twin of Eq.~(\ref{y1}). For a
given observed period $T$, Eq.~(\ref{zeta1}) is a parametric
representation of $\zeta$ in terms of $y$ (or equivalently in terms
of $\tau_{\rm A,i}$). In order to stress this point we define the
function $G_1$ by use of the right hand member of Eq.~(\ref{zeta1}) as
\begin{equation}
G_1 \;:\;[\frac{\displaystyle 1}{\displaystyle 2}, \;\;
\frac{\displaystyle 1}{\displaystyle \sqrt{2}} [ \;\;\rightarrow
\mathbf{R},\;\; y \; \Rightarrow \; G_1(y) = \frac{\displaystyle 2
y^2}{\displaystyle 1 -  2 y^2}. \label{G1}
\end{equation}
Since ${\mathrm d}G_1/{\mathrm d} y  > 0 $
and ${\mathrm d}^2 G_1 / {\mathrm d} y^2 > 0$, it follows that  $\zeta=G_1(y)$ is a strictly increasing
and convex function of $y$. 
Note now that 
$$
G_1(1/2) = 1, \;\;\lim_{y \rightarrow 1/\sqrt{2}} G_1(y)= \infty.
$$
With this information on the function $G_1$ we can refine its
definition (\ref{G1}). Combined with the definition (\ref{F1a}) of
$F_1$ we obtain the following prescriptions for the functions $F_1$
and $G_1$;

\begin{eqnarray}
F_1 \;&:&\;[1, \; \infty[ \;\rightarrow  [\frac{\displaystyle
1}{\displaystyle 2}, \;\; \frac{\displaystyle 1}{\displaystyle
\sqrt{2}} [,\;\; \zeta \; \Rightarrow \; F_1(\zeta) =
\frac{\displaystyle 1}{\displaystyle \sqrt{2}} \left(
\frac{\displaystyle \zeta}{\displaystyle \zeta + 1} \right )^{1/2},
\nonumber \\
G_1 \;&:& \;[\frac{\displaystyle 1}{\displaystyle 2}, \;\;
\frac{\displaystyle 1}{\displaystyle \sqrt{2}} [ \;\;\rightarrow [1,
\; \infty[,\;\; y \; \Rightarrow \; G_1(y) = \frac{\displaystyle 2
y^2}{\displaystyle 1 -  2 y^2}. \label{F1aG1a}
\end{eqnarray}

Let us recapitulate what seismic information we have deduced from
the observed value of the period. The quantities $\zeta$ and $y$ are
in the following intervals
\begin{equation}
\zeta \in I_{\zeta} = [1, \;\infty[, \;\;\;y \in I_y =
[\frac{\displaystyle 1}{\displaystyle 2}, \;\; \frac{\displaystyle
1}{\displaystyle \sqrt{2}} [.
\end{equation}
and are related to one another by
\begin{equation}
y = F_1(\zeta), \;\;\;\zeta = G_1(y). \label{yzeta}
\end{equation}
The functions $F_1$ and $G_1$ are defined by expressions~(\ref{F1aG1a}). Of course, $G_1$ 
is the inverse function of $F_1$: $G_1 =
F_1^{-1}$ and conversely $G_1^{-1} = F_1$.  When
the period $T$ is known from observations, then there are
infinitely many pairs $(\zeta, y)$  that reproduce the observed
period. We can let $\zeta$ vary over the interval $[1,
\;\infty[$ and compute for each value of $\zeta$ the
corresponding value of $y = F_1(\zeta)$, or conversely, let $y$
vary over the interval $[ 1/2, \; 1 /\sqrt{2} [$ and compute for
each value of $y$ the corresponding value of $\zeta = G_1(y)$.

\section{Asymptotic analytic expressions for damping time}

In order for the kink MHD waves to be damped by resonant absorption additional physics has
to be introduced in the equilibrium model. The required additional
physics is non-uniformity of the local Alfv\'{e}n velocity. For a
constant magnetic field this implies a non-uniform density.
Asymptotic expressions for the damping time have been derived by
e.g. \cite{HY88}, \cite{Goossens92} and
\cite{RR02}. These asymptotic expressions are derived in
the approximation that the non-uniform layer is thin.  
The true density
discontinuity is replaced by a continuous variation in density. 
Jump conditions are used to connect the solution over the ideal
singularity and to avoid solving the non-ideal MHD wave equations.
The jump condition for the ideal Alfv\'{e}n singularity was
introduced on an intuitive manner by \cite{HY88} and put on
a firm mathematical basis by \cite{sakurai91},
\cite{Goossens95} and \cite{GORU95} for the driven
problem, and by \cite{TIGO96} for the eigenvalue problem. The
result of this asymptotic analysis is
\begin{equation}
\frac{\displaystyle \tau_{\rm d}}{\displaystyle T} = F  \;\; \frac
{\displaystyle R} {\displaystyle l}\;\; \frac{\displaystyle \rho_{\rm i} +
\rho_{\rm e}}{\displaystyle \rho_{\rm i} - \rho_{\rm e}}.  \label{taud}
\end{equation}
Here $T$ and $\tau_{\rm d}$ are the period and the damping time,  $\rho_{\rm i}$
is the internal density on the axis of the loop and hence in the
interval $[0,\; R - \frac{l}{2}]$; $\rho_e$ is the constant
external density where external refers to $[R + \frac{l}{2}, \;
\infty[$, $l$ is the thickness of the non-uniform layer; $l/R =0$
corresponds to a uniform loop and a discontinuous variation in the
density at the radius $R$ of the loop.  A fully non-uniform loop has
$l/R = 2$ but cannot be studied by an approximate TB expression
for the damping time. The numerical factor $F$ depends on
the variation in density across the non-uniform layer. For a
linear variation as used by \cite{Goossens92} $F = 4/\pi^2$.
\cite{RR02} used a sinusoidal variation in density across
the non-uniform layer and found $F = 2/\pi$. In what follows we
shall adopt $F = 2/\pi$.  A sinusoidal variation in density in the non-uniform
transitional layer is probably closer to reality than a linear
variation. In addition we shall compare our analytic results with 
numerical results for fully non-uniform 1-dimensional models of coronal
loops obtained by \cite{tom04b} and by
\cite{Arregui07} by the use of the eigenvalue code LEDA
originally designed by \cite{vdl91}. In these studies a sinusoidal
variation in density was considered. Note also that in Eq.~(\ref{taud}) the period is
computed using the TT-approximation.

Rewrite Eq.~(\ref{taud}) in terms of $\zeta$ to find
\begin{equation}
\frac{\displaystyle \tau_{\mathrm d}}{\displaystyle T} = \frac
{\displaystyle 2} {\displaystyle \pi} \frac{\displaystyle \zeta +
1}{\displaystyle \zeta - 1} \frac{\displaystyle 1}{\displaystyle
l/R}. \label{DecayT}
\end{equation}

Equation~(\ref{DecayT}) is the second key equation of the present
investigation. Let us now look at this equation from a seismic
point of view. If we have observed values  of the period $T$ and
the damping time $\tau_{\rm d}$ and we convince ourselves that
Eq.~(\ref{DecayT}) is a good first analytical approximation of the
damping time divided by period then we can invert Eq.~(\ref{DecayT})
for either $\zeta$ or $l/R$. Actually we shall do both. Let us
first solve Eq.~(\ref{DecayT}) for $\zeta$ and find
\begin{equation}
\zeta = \frac{\displaystyle \frac{\displaystyle l}{\displaystyle 2
R}\;\; \frac{\displaystyle \pi \tau_{\rm d}}{\displaystyle T} + \;1} {
\frac{\displaystyle l}{\displaystyle 2 R} \;\;\frac{\displaystyle
\pi \tau_{\rm d}}{\displaystyle T} - 1}.\label{zeta2}
\end{equation}
From a seismic point of view, period and damping time can be considered as known
from observations and it makes sense to denote $\pi \tau_{\rm d}/ T$ as a
constant $C$. In addition it is convenient to abbreviate $l/(2R)$ as
$z$. Hence
\begin{equation}
z = \frac{\displaystyle l}{\displaystyle 2 R}, \;\;C
=\frac{\displaystyle \pi \tau_{\rm d}}{\displaystyle T}. \label{z}
\end{equation}
Equation~(\ref{zeta2}) can be rewritten as
\begin{equation}
\zeta = \frac{\displaystyle C z +1}{\displaystyle C z -1}.
\label{zeta3}
\end{equation}
Here are several observations to be made. First, since $\zeta >0$ it
follows from Eq.~(\ref{zeta3}) that for a given ratio of $\tau_{\rm d}/T$
there is a lower  bound for the inhomogeneity length scale

\begin{equation}
z = \frac{\displaystyle l}{\displaystyle 2 R} > \frac{\displaystyle
T}{\displaystyle \pi \tau_{\rm d}} = \frac{\displaystyle 1}{\displaystyle
C} = z_{\rm min}. \label{zmin}
\end{equation}
Second, realize that Eq.~(\ref{zeta3}) is a parametric representation of
$\zeta$ in terms of $z= l/(2R)$. In order to make this point very
explicit, we introduce the function $G_2$ defined as
\begin{equation}
G_2\;:\;]\frac{\displaystyle 1}{\displaystyle C}, \;\;\; 1]
\;\rightarrow \mathbf{R},\;\; z \; \Rightarrow \; G_2(z) =
\frac{\displaystyle C z + 1}{\displaystyle C z - 1}. \label{G2}
\end{equation}
It is easy to show that ${\mathrm d} G_2/{\mathrm d} z <0$ and ${\rm d}^2 G_2 /{\rm d} z^2 > 0$.
Hence $\zeta=G_2(z)$ is a strictly decreasing and convex function of $z$. It attains
its absolute minimum for $z = l/(2R) = 1$. This minimal value is
\begin{equation}
\zeta_{\rm min} = G_2(1) =  \frac{\displaystyle C +
1}{\displaystyle C - 1}.
 \label{zetamin}
\end{equation}
Conversely $\zeta$ attains its maximal value of infinity in the limit $z
\rightarrow 1/C$.
With this information on the function $G_2$ we can refine its
definition (\ref{G2}) as follows

\begin{equation}
G_2\;:\;]\frac{\displaystyle 1}{\displaystyle C}, \;\;\; 1]
\;\rightarrow\;[\frac{\displaystyle C + 1}{\displaystyle C - 1},
\;\;\infty[,\;\; z \; \Rightarrow \; G_2(z) = \frac{\displaystyle
C z + 1}{\displaystyle C z - 1}. \label{G2a}
\end{equation}

With the minimal value  (\ref{zetamin}) for $\zeta$ we can slightly
improve on the lower bound $1/2$ for $y$ and $T/2$ for $\tau_{\rm A,i}$
given by inequality~(\ref{y2}) and on the upper bound $2 L/T$ for $V_{\rm A,i}$
given by inequality~(\ref{vAi}).  These bounds were found with the sole use of
the information on the periods. If we also use the information on
the damping rate then we obtain
\begin{equation}
y \geq \frac{\displaystyle 1}{\displaystyle 2} \left
(\frac{\displaystyle C + 1}{\displaystyle C} \right )^{1/2},\;
\tau_{\rm A,i} \geq \frac{\displaystyle T}{\displaystyle 2} \left
(\frac{\displaystyle C + 1}{\displaystyle C} \right )^{1/2}, \;\;
V_{\rm A,i} \leq \frac{\displaystyle 2 L }{\displaystyle T} \left
(\frac{\displaystyle C}{\displaystyle C + 1} \right )^{1/2}.
\label{vAi2}
\end{equation}

With the help of the information on the bounds for $y$ and $\zeta$
we can refine the definitions (\ref{F1aG1a}) for $F_1$ and $G_1$
respectively to their  final versions
\begin{eqnarray}
F_1 \;&:&\;[\frac{\displaystyle C + 1}{\displaystyle C - 1},
\;\;\infty[ \;\rightarrow [\frac{\displaystyle 1}{\displaystyle 2}
\left (\frac{\displaystyle C + 1}{\displaystyle C} \right )^{1/2} ,
\;\; \frac{\displaystyle 1}{\displaystyle \sqrt{2}}[ ,\;\; \nonumber\\ 
&&\zeta \; \Rightarrow \; F_1(\zeta) = \frac{\displaystyle 1}{\displaystyle
\sqrt{2}} \left( \frac{\displaystyle \zeta}{\displaystyle \zeta + 1}
\right )^{1/2},  \nonumber\\
&& \mbox{} \nonumber\\
G_1 \;&:&\; [\frac{\displaystyle 1}{\displaystyle 2} \left
(\frac{\displaystyle C + 1}{\displaystyle C} \right )^{1/2} , \;\;
\frac{\displaystyle 1}{\displaystyle \sqrt{2}}[\;\;\rightarrow
[\frac{\displaystyle C + 1}{\displaystyle C - 1}, \;\;\infty[ ,\;\; \nonumber\\
&&
y \; \Rightarrow \; G_1(y) = \frac{\displaystyle 2
y^2}{\displaystyle 1 -  2 y^2}. \label{F1G1}
\end{eqnarray}

Let us now solve Eq.~(\ref{DecayT}) for $z= l/(2R)$ and find
\begin{equation}
z = \frac{\displaystyle l}{\displaystyle 2 R} = \frac{\displaystyle
1} {\displaystyle C} \frac{\displaystyle \zeta + 1}{\displaystyle
\zeta - 1}. \label{z2}
\end{equation}
Equation~(\ref{z2}) is a parametric representation of $z = l/(2R)$ in terms
of $\zeta$. As before we make this point explicit by introducing the
function $F_2$ defined as

\begin{equation}
F_2 \;:\;[\frac{\displaystyle C + 1}{\displaystyle C - 1}, \;\;\;
\infty [ \;\rightarrow \mathbf{R},\;\; \zeta \; \Rightarrow \;
F_2(\zeta) =  \frac{\displaystyle 1}{\displaystyle C}
\frac{\displaystyle \zeta + 1}{\displaystyle \zeta - 1}.
\label{F2}
\end{equation}
It is easy to show that ${\mathrm d}F_2/{\mathrm d}\zeta  <0$ and ${\rm d}^2 F_2 / {\rm d} \zeta^2 > 0$, hence
$z=F_2(\zeta)$ is a strictly decreasing and convex function of $\zeta$. In addition
$$
F_2(\zeta_{\rm min}) = 1, \;\;\lim_{\zeta \rightarrow \infty}
F_2(\zeta) = \frac{1}{C}.
$$
With this information on the function $F_2$ we can refine its
definition (\ref{F2}). Combined with the definition (\ref{G2a}) for
the function $G_2$ we obtain

\begin{eqnarray}
F_2 \;&:& \;[\frac{\displaystyle C + 1}{\displaystyle C - 1},
\;\;\; \infty [ \;\rightarrow ]\frac{\displaystyle
1}{\displaystyle C}, \;\;\; 1],\;\; \zeta \; \Rightarrow \;
F_2(\zeta) = \frac{\displaystyle 1}{\displaystyle C}
\frac{\displaystyle \zeta + 1}{\displaystyle \zeta - 1}, \nonumber \\
&& \mbox{} \nonumber \\
G_2\;&:&\;]\frac{\displaystyle 1}{\displaystyle C}, \;\;\; 1]
\;\rightarrow\;[\frac{\displaystyle C + 1}{\displaystyle C - 1},
\;\;\infty[,\;\; z \; \Rightarrow \; G_2(z) = \frac{\displaystyle
C\; z + 1}{\displaystyle C\; z - 1}. \label{F2G2}
\end{eqnarray}
Here $F_2$ is the inverse function of $G_2$; $F_2 = G_2^{-1}$ and
conversely $G_2$ is the inverse function of $F_2$.

\begin{table*}[t]
\caption{{\em Left}: Loop oscillation properties of the analysed events. {\em Right}: Analytic (A) and numerical (N) inversion results.}
\label{restable}
\centering
\begin{tabular}{c c c c c c c|c cccc}
\hline\hline
Loop  & R & L & R/L & T & $\tau_{\rm d}$ & $T/\tau_{\rm d}$& $\tau_{\rm A,iA}$ & $\tau_{\rm A,iN}$& $V_{\rm A,iN}=L/\tau_{\rm A,iN}$ \\
& (10$^{6}$ m)& (10$^{8}$ m) & (10$^{-2}$) & (s) & (s) & & (s)& (s)& (km s$^{-1}$)&\\
\hline
1 &  3.60 &  1.68 &  2.13 &  261 &  870 &  0.30  &137 (143)--185& 145--177 & 947--1161&\\

2 &  3.35 &  0.72 &  4.65 &  265 &  300 &  0.88 &150--187&  163--182  & 396--443&\\

3 &  4.15 &  1.74 &  2.37 &  316 &  500 &  0.63 & 173--224& 189--217  & 801--922&\\

4 &  3.95 &  2.04 &  1.94 &  277 &  400 &  0.69 & 153--196& 168--189  & 1079--1211&\\

5 &  3.65 &  1.62 &  2.26 &  272 &  849 &  0.32 &143 (150)--192&  151--187  & 869--1073&\\

6 &  8.40 &  3.90 &  2.17 &  522 & 1200 & 0.44 & 279 (286)--369& 304--359 & 1088--1284&\\

7 &  3.50 &  2.58 &  1.37 &  435 &  600 &  0.73& 241--308& 267--299  & 862--967&\\

8 &  3.15 &  1.66 &  1.88 &  143 &  200 &  0.72& 79--101&  90--98  & 1693--1853&\\

9 &  4.60 &  4.06 &  1.15 &  423 &  800  & 0.53& 229 (232)--299 &247--291 &   1394--1643&\\

10 & 3.45 &  1.92 & 1.78 & 185  &  200 &  0.93 & 105--131&117--126 & 1526--1643&\\

11 &7.90  & 1.46 &  5.38 & 390 &  400 &  0.98 & 227--280& 245--270   & 541--595&\\
\hline
\end{tabular}

\end{table*}

\section{Analytical seismology and comparison with numerical inversion}\label{seismology}
Let us recapitulate the key results of the previous section. The
two quantities that we assume to be known from observations are
the period $T$ and the damping time $\tau_{\rm d}$. Analytical theory
based on the TTTB-approximation gives us two equations, namely
Eqs.~(\ref{T}) and (\ref{DecayT}) that express the period $T$ and the
damping time $\tau_{\rm d}$ in terms of the density contrast $\zeta$,
the Alfv\'{e}n transit time (normalised to the period) $y =
\tau_{\rm A,i}/T$ and the inhomogeneity length scale (normalised to
the radius of the loop) $z = l/(2R)$. These three quantities
$\zeta, y$ and $z$ are the seismic quantities in the sense that
they are the quantities that we aim to determine with the use of
observed values of the period $T$ and the damping time $\tau_{\rm d}$.
Since we have only two equations that relate the three unknown
quantities to the two observed quantities there are an infinite
number of solutions of the seismic inversion problem, such as 
first pointed out by \cite{Arregui07}. The seismic
variables are constrained to the following intervals
\begin{eqnarray}
\zeta &\; \in \; & I_{\zeta} = [\frac{\displaystyle C +
1}{\displaystyle C - 1}, \;\;\infty[ \nonumber \\
y & \; \in \; & I_y = [\frac{\displaystyle 1}{\displaystyle 2} \left
(\frac{\displaystyle C + 1}{\displaystyle C} \right )^{1/2} , \;\;
\frac{\displaystyle 1}{\displaystyle \sqrt{2}}[ \nonumber \\
z & \;\in \; & I_z = ]\frac{\displaystyle 1}{\displaystyle C},
\;\;1]
\label{yzez1}
\end{eqnarray} 
and are related to one another by
\begin{eqnarray}
y &\; =\; & F_1(\zeta), \;\;\; \zeta \; =\;  G_1(y), \nonumber \\
z & \; =\; & F_2(\zeta),\;\;\; \zeta  \; =\; G_2(z). \label{yzez2}
\end{eqnarray}
The functions $F_1, G_1, F_2, G_2$ are defined by (\ref{F1G1}) and
(\ref{F2G2}).

Of the four functions only two  are
independent since $G_1$ is the inverse function of $F_1$ and $G_2$
is the inverse function of $F_2$. The set (\ref{yzez2}) gives us the
infinitely many solutions of the seismic inversion in parametric
form. Each of the three unknowns can be used as parameter and the
two remaining unknowns can be expressed in terms of that
parameter. For example choose $\zeta$ as parameter. Let $\zeta$
take on all values in $I_{\zeta}$ and compute the corresponding
values of $y$ and $z$ by the use of $y = F_1(\zeta)$ and $z =
F_2(\zeta)$. Or choose $y$ as parameter. Let $y$ take on all
values in $I_y$ and then compute the corresponding values of
$\zeta$ and $z$ by the use of $\zeta = G_1(y)$ and $z =
F_2(G_1(y)).$ Finally, use $z$ as parameter to define the
solutions of the inversion problem. Let $z$ take on all values in
$I_z$ and then compute the corresponding values of $y$ and $\zeta$
by the use of $\zeta = G_2(z)$ and $y = F_1(G_2(z))$.

As an illustrative example we re-analyse loop oscillation event \#
5 examined by \cite{Arregui07} in detail using their numerical
seismic inversion scheme. The results of the investigation of that
loop event are shown on their Figure 3a. For this loop oscillation
event $T = 272 \;\mbox{s}$ and $\tau_{\rm d} = 849 \;\mbox{s}$. The
radius $R$ and the length $L$ of this loop are estimated to be $R=
3.65 \;\times 10^6 \; \mbox{m}$ and $L = 1.62 \times 10^8 \;\mbox{m}$
respectively. The ratio of the radius to the length of the loop is
$R/L \approx 2 \times 10^{-2}.$ This small value is good news for
the TT-approximation to the period with effects due to a non-zero
radius on period being small.

The constant $C$=$(\pi\tau_{\rm d})/T$ = $9.81$.  Hence
$\zeta_{\rm{min}} = 1.23, \; y_{\rm{min}} = 0.525, \;
y_{\rm{max}} = 0.707$ and $ z_{\rm{min}} =0.102$. The
intervals for $\zeta, y, z$ are now
\begin{eqnarray}
\zeta &\; \in \; & I_{\zeta} = [1.23, \;\;\infty[, \nonumber \\
y & \; \in \; & I_y = [0.525 , \;\;0.707[, \nonumber \\
z & \;\in \; & I_z = \;]0.102, \;\;1].
\end{eqnarray} \label{yzetaz2}
The corresponding interval for $\tau_{\rm A,i}$ is
$$
143 \;\mbox{s} \;\leq \;\tau_{\rm A,i} \;\leq \;192 \;\mbox{s}.
$$
If we limit the analysis to $\zeta \geq 1.5$ then the lower bound is
$ 150 \;\mbox{s}$ so that
$$
150 \;\mbox{s} \;\leq \;\tau_{\rm A,i} \;\leq \;192 \;\mbox{s}.
$$
This can be compared with the interval in Fig 3a of
\cite{Arregui07} which is (see also their Table 1)
$$
170 \;\mbox{s} \;\leq \;\tau_{\rm A,i} \;\leq \;210  \;\mbox{s}.
$$

It is encouraging to note that the overall differences on the
Alfv\'{e}n travel times found here are about $10\%$ although the
loop oscillation event is characterised by heavy damping. It is
intriguing that the interval as a whole is shifted by about
$20\;\mbox{s}$ to shorter Alfv\'{e}n travel times.
This intriguing discrepancy has led us to calculate  analytical results for the allowable range 
of the Alfv\'en travel time for the remaining 10 loops and to compare them with those obtained 
by \citet{Arregui07}. It turned out that in all cases there are discrepancies and
some of them were far bigger than 10$\%$. 
This has motivated us to re-examine the numerical results of the investigation of \citet{Arregui07}.
It turns out that the values given in \citet{Arregui07} are inaccurate. 
We have re-calculated  the numerical values for the Alfv\'en travel time intervals in 
\citet{Arregui07} and present them in Table~\ref{restable}. 
When this issue is taken into account, the corrected numerical interval  for the event under consideration is
$$
151 \;\mbox{s} \;\leq \;\tau_{\rm A,i} \;\leq \;187 \;\mbox{s},
$$
which remarkably agrees with the analytic interval. 
Table~\ref{restable} shows that analytic and numerical Alfv\'en travel time intervals
agree very well for all events. Analytic lower bounds for $\tau_{\rm A,i}$ are a little below numerical ones. 
Note that analytic lower bounds for $\zeta$ given in (\ref{yzez1}) can be slightly lower or larger than 
the numerical  $\zeta=1.5$. When lower, values in parentheses give the analytic $\tau_{\rm A,i}$ for this density contrast. 
Upper bounds correspond to the limit $\tau_{\rm A,i}(\zeta\rightarrow\infty)=T/\sqrt{2}$, which is 
nearly approximated at the numerical $\zeta\simeq20$. We have also calculated $z_{\rm min} = 1/C$  for 
the events studied by \citet{Arregui07} and the results agree well with the corresponding 
values of $(l/R)_{\rm min}$ in their Table 1.

\begin{table}[t]
\caption{Analytic seismic inversion results for loop \# 5.}
\label{table2}
\centering
\begin{tabular}{lcc}
\hline\hline
$z=l/(2R)$ &  $\zeta$ & $y=\tau_{\rm Ai}/T$  \\
\hline

  0.105  & 67.67  &  0.702  \\
  0.110  & 26.32  &  0.694  \\
  0.120  & 12.30  &  0.680  \\
  0.125  &  9.85  &  0.674  \\
  0.150  &  5.24  &  0.648  \\
  0.175  &  3.79  &  0.629  \\
  0.200  &  3.08  &  0.617  \\
  0.225  &  2.66  &  0.603  \\
  0.250  &  2.38  &  0.593  \\
  0.275  &  2.18  &  0.585  \\
  0.300  &  2.03  &  0.579  \\
  0.325  &  1.91  &  0.573  \\
  0.350  &  1.82  &  0.568  \\
  0.375  &  1.75  &  0.564  \\
  0.400  &  1.68  &  0.560  \\
  0.425  &  1.63  &  0.557  \\
  0.450  &  1.59  &  0.554  \\
  0.475  &  1.55  &  0.551  \\
  0.500  &  1.51  &  0.549  \\
  0.525  &  1.48  &  0.546  \\
\hline
\end{tabular}
\end{table}

As an illustration that any of the three seismic quantities can be
used as parameter we take $z$ as the parameter. We let $z$ vary
and compute for each value of $z$ the
corresponding value of $\zeta$ by the use of the function $G_2$
defined in (\ref{F2G2}),
\begin{equation}
\zeta = \frac{\displaystyle 9.81 z + 1}{\displaystyle 9.81 z - 1}.
\label{zeta5}
\end{equation}
Subsequently we compute the corresponding value of $y$ by the use
of the function $F_1$ defined in (\ref{F1G1}),
\begin{equation}
y = \frac{\displaystyle 1}{\displaystyle \sqrt{2}} \left(
\frac{\displaystyle \zeta}{\displaystyle \zeta + 1} \right )^{1/2}.
\label{y5}
\end{equation}

\begin{figure}[!t]
   \resizebox{\hsize}{!}{\includegraphics{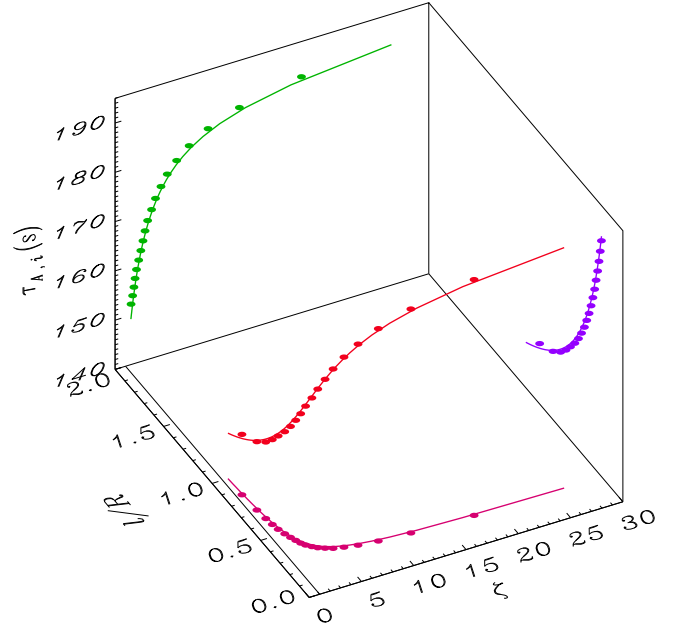}}
    \caption{Analytic inversion (solid lines) and corrected numerical inversion (filled circles) in the 
    ($\zeta$, $l/R$, $\tau_{\rm A,i}$)-space for loop oscillation event \#5 in Table~\ref{restable}.}
    \label{fig1}
   \end{figure}

Of course, we can only compute $\zeta$ and $y$ for discrete values
of $z$. The results of our computations
are summarised on Table~\ref{table2} and graphically represented on Fig.~\ref{fig1}.
Recall that the functions  $F_1, G_1, F_2, G_2$ (see eqs.
[\ref{F1G1}] and [\ref{F2G2}]) are monotonically increasing ($F_1,
G_1$) or monotonically decreasing ($F_2, G_2$) and have concave
graphs ($F_1$) and convex graphs ($F_2, G_1, G_2$) respectively.
This implies f.e. that $y = \tau_{\rm A,i}/T$ is a strictly increasing
function of $\zeta$ and a strictly decreasing function of $z$ and
conversely that $z$ is a strictly decreasing function of both
$\zeta$ and $\tau_{\rm A,i}$. 
Inspection of the second order
derivative of a given quantity with respect to one of the two
remaining quantities can tell us that the graph is either concave
or convex. 
For example the graphs of $y$ and of $z$ as function of
$\zeta$ are respectively concave and convex. The monotonic
variation of $\zeta, y, z$ and the concave or convex appearance of
their graphs predicted by our analytic seismic inversion agrees
exactly with the behaviour of the numerical inversion. Furthermore, 
Fig.~\ref{fig1} displays an amazing quantitative agreement between analytic 
and numerical inversion results.

We have not re-analysed in detail loop oscillation event \# 10 that was
examined by \cite{Arregui07}. The inversion for that
loop event is shown on their Figure 3b. A striking property of the
solutions is the non-monotonic behaviour of the seismic variables.
This is clearly reflected in the pronounced minimum of
$\tau_{\rm A,i}$ as function of $\zeta$ and as function of $l/R$. The
decreasing part of $\tau_{\rm A,i}$ as function of $\zeta$ and the
increasing part of $\tau_{\rm A,i}$ as function of $l/R$  cannot be
recovered by the analytical seismic inversion scheme based on the
TTTB-approximation. The analytical TTTB-approximation predicts
monotonic variation of the seismic variables and cannot
approximate multi-valued solutions with two pairs of $(\zeta,
l/R$) corresponding to the same value of $\tau_{\rm A,i}$. The fact
that the analytical seismic inversion fails for this loop
oscillation event is not disturbing since this event is
characterised by extremely heavy damping with $T/\tau_{\rm d} = 0.92$
which we anticipated would fall out of the application range of
the analytical scheme anyway.

\section{Conclusion} In this paper we have presented an analytic
approximate seismic inversion scheme based on the
TTTB-approximation for computing the period and the damping time.
In the TTTB-approximation the period is computed for a uniform
loop model in the long wavelength or zero radius approximation.
The damping time is computed for relatively weak damping
corresponding to thin non-uniform layers. The advantage of this
analytical seismic inversion is that it is formulated with the aid
of two functions $F_1$ and $F_2$ (and their inverse functions
$G_1 = F_1^{-1}$ and $G_2 = F_2^{-1}$) which  are given by simple
closed expressions. The practical implementation of the inversion
scheme is stunningly simple. The calculations required to obtain
solutions can even done with the use of a hand calculator. This
analytical scheme seismic inversion clearly shows that the
inversion problem has infinitely solutions in the $(\zeta, y,
z)$-space as first pointed out by \cite{Arregui07}. It also
reveals that the allowable values of $y$ (or Alfv\'{e}n travel
time) are confined to a narrow range. When applied to a loop
oscillation event with heavy damping as f.e. loop oscillation
event \# 5 with $T/\tau_{\rm d} = 0.32$ the analytic inversion scheme
produces remarkably accurate results. Not only does it recover the
overall appearance of the solution curve with the corresponding
monotonic behaviour of the seismic variables. In addition, it
recovers for a prescribed range of values of $\zeta$ the
corresponding values of $y$ (or $\tau_{\rm A,i}$) and $z$ (or $l/R$).

The disadvantage of the scheme is that (i) it does not take into
account the effects of non-zero radius and of radial
stratification on the period; (ii) it is strictly speaking only
valid for weak damping corresponding to thin non-uniform layers. 
Corrections due to finite tube radius are of the order of the loop radius to length
ratio squared. For the largest observed value of $R/L$ in Table~\ref{restable} the 
correction is of the order of $10^{-3}$, hence the thin tube approximation does 
not impose any practical restriction on the applicability of analytical results.
As for the thin boundary approximation, for cases with extremely 
heavy damping, the non-monotonic behaviour displayed by numerical solutions cannot be recovered.
This is the price for an analytical scheme. All in all, the accuracy of the results
obtained with this analytic inversion is amazing.
The final agreement of the analytic seismic inversion with the numerical seismic inversion when 
the inaccuracies of the numerical inversion are removed is excellent up to the 
point that the analytic seismic inversion emerges as a tool for validating results of numerical 
inversions.

\begin{acknowledgements}
This research was begun when MG was a visitor of the Solar Physics
Group at the UIB.  It is pleasure for MG to acknowledge the warm
hospitality of the Solar Physics Group at the UIB  and the visiting
position from the UIB. MG also acknowledges
the  FWO-Vlaanderen for awarding him a sabbatical leave.
IA and JLB acknowledge the funding provided under projects AYA2006-07637 (Spanish MEC)
and PRIB-2004-10145 and PCTIB2005GC3-03 (Government of the Balearic Islands).
TJW's work was supported by NRL grant N00173-06-1-G033.
\end{acknowledgements}


\end{document}